\documentclass[letterpaper,10pt,conference]{IEEEtran} 

\IEEEoverridecommandlockouts                              
\usepackage{graphics} 
\usepackage{epsfig} 
\usepackage{mathptmx} 
\usepackage{amsmath} 
\usepackage{amssymb}  
\usepackage[table,xcdraw]{xcolor} 
\usepackage{array}
\usepackage{tabularx} 
\usepackage{booktabs} 
\usepackage{multirow} 
\usepackage{ragged2e} 
\usepackage{hhline} 
\usepackage{cellspace} 
\usepackage{titlesec} 
\usepackage{graphicx} 
\usepackage{enumitem} 
\usepackage{nicematrix} 
\usepackage{float} 
\usepackage{hyperref} 
\usepackage[table]{xcolor}
\usepackage[absolute,overlay]{textpos} 

\title{\LARGE \bf
On the Standard Performance Criteria for Applied Control Design: PID, MPC or Machine Learning Controller?}

\author{Pouria Sarhadi$^\dag$
\thanks{$^\dag$Pouria Sarhadi is with the School of Physics, Engineering and Computer Science, University of Hertfordshire, Hatfield, UK        ({\tt\small p.sarhadi@herts.ac.uk)}}}

\definecolor{LR}{rgb}{1.0, 0.5, 0.5}  
\definecolor{DR}{rgb}{0.97, 0.0, 0.0}  
\definecolor{LG}{rgb}{0.5, 1.0, 0.5}  
\definecolor{DG}{rgb}{0.0, 0.8, 0.0}  
\definecolor{WH}{HTML}{FFFFFF}  
\definecolor{LB}{HTML}{E0F0FF}  
\definecolor{MB}{HTML}{C0D9FF}  
\definecolor{darkgreen}{rgb}{0.0, 0.5, 0.0}  

\begin{document}

\maketitle
\thispagestyle{empty}
\pagestyle{empty}

\begin{abstract}
The traditional control theory and its application to basic and complex systems have reached an advanced level of maturity. This includes aerial, marine, and ground vehicles, as well as robotics, chemical, transportation, and electrical systems widely used in our daily lives. The emerging era of data-driven methods, Large Language Models (LLMs), and AI-based controllers does not indicate a weakness in well-established control theory. Instead, it aims to reduce dependence on models and uncertainties, address increasingly complex systems, and potentially achieve decision-making capabilities comparable to human-level performance. This revolution integrates knowledge from computer science, machine learning, biology, and classical control, producing promising algorithms that are yet to demonstrate widespread real-world applicability. Despite the maturity of control theory and the presence of various performance criteria, there is still a lack of standardised metrics for testing, evaluation, Verification and Validation (V\&V) of algorithms. This gap can lead to algorithms that, while optimal in certain aspects, may fall short of practical implementation, sparking debates within the literature. For a controller to succeed in real-world applications, it must satisfy three key categories of performance metrics: tracking quality, control effort (energy consumption), and robustness.
This paper rather takes an applied perspective, proposing and consolidating standard performance criteria for testing and analysing control systems, intended for researchers and students. The proposed framework ensures the post-design applicability of a black-box algorithm, aligning with modern data analysis and V\&V perspectives to prevent resource allocation to systems with limited impact or imprecise claims.
\end{abstract}

\section{INTRODUCTION}
\begin{textblock*}{3.4in}(0.75in, 10.22in) 
    \footnotesize This work has been submitted to the IEEE for possible publication. Copyright may be transferred without notice, after which this version may no longer be accessible.
\end{textblock*}
Over a century of remarkable theoretical and applied development, we have observed the widespread exploitation of control in various systems. While myriads of control algorithms have been developed, (advanced) Proportional-Integral-Derivative (PID) controllers with essential modifications \cite{skogestad2023advanced} and Model Predictive Control (MPC) remain the most widely used in industrial applications \cite{hagglund2024PIDMPC}. These techniques have proven successful in controlling everyday real systems, with their performance approaching a maximum saturation point. However, relentless ambition continues to drive the quest for controllers with human-level performance. This pursuit has led to new endeavours in control design, ranging from data-driven methods to Machine Learning (ML)-based approaches, not merely to achieve marginal performance improvements but to create more intelligent systems with reduced need for redesign and retuning, thereby enhancing sustainability in design. As a result, advanced control algorithms with notable attention (around 20k citations in 2025, possibly more than any control engineering paper in the recent decade) have emerged \cite{lillicrap2015continuousddpg}, bridging computer science and control engineering. Yet, they still require a path to establish trust for real-world applications. Although the premise of this paper may seem familiar, it offers a rich applied control perspective and addresses often disregarded aspects in new developments. While various control design evaluation metrics are known, the control community still lacks explicitly defined benchmark criteria and testing methodologies. It is shown that, despite promotions and discussions in the literature, some basic methods can demonstrate better or more reliable performance \cite{skogestad2023advanced,skogestad2019smith,narendra2024ML}. One reason for this is the absence of rigorous testing frameworks and problem definitions from an applied control perspective to ensure that newly developed approaches are viable for implementation. This need is particularly pressing given the vast number of publications in the field, many of which focus on certain aspects of algorithmic optimality while overlooking critical challenges in applied control design. This may stem either from the intense focus of the control community on theoretical developments, often restricted to aspects such as error signal convergence, or from the reliance of the computer science community on coding capabilities and conceptual innovation, sometimes neglecting operational challenges. In any case, the ultimate output is a controller that must be tested against real-world problems and quantitatively assessed through rigorous testing criteria. This paper aims to introduce a set of standardised performance criteria for evaluating, testing, and analysing controllers to demonstrate algorithm applicability.
\setlength{\unitlength}{1in}

The analysis of control system performance dates back to the mid-20th century, with key results reviewed in 1961 \cite{ControlPerformance1961}. Some insightful analyses are presented in \cite{ControlPerformance1961}, but they primarily focus on the error signal. The Control Handbook \cite{levine2018control} has emphasised similar error-based criteria, such as ISE and ITAE (discussed later), which are essential but not sufficient. The emergence of robust control, sensitivity analysis \cite{skogestad2005multivariable}, and historical developments in adaptive controllers—such as the stability challenges of the NASA's X-15 crash \cite{dydek2010adaptiveX15} and Rohrs' example \cite{rohrs1985robustness}—highlighted the need for rigorous robustness metrics and, more importantly, well-defined testing methodologies. However, these have not been standardised. One of the best contemporary examples is the European Cooperation on Space Standardisation's (ECSS) control performance guideline, which considers tracking metrics alongside robustness measures such as gain, phase, and delay margins \cite{ECSS2010}. The idea of defining benchmark problems in control, like the 1992 example \cite{wie1992benchmark}, was excellent but loosely continued and now seems to be regaining interest \cite{maestre2025benchmark2}. It has long been recognised that tracking and robustness alone are insufficient; neglecting control effort and the consequences of amplitude and rate saturation can lead to catastrophic failures, such as the Chernobyl disaster or various aircraft crashes \cite{SteinRespect}. While the importance of control effort has been acknowledged in optimal controllers like Linear Quadratic Regulators (LQR) and MPC through cost functions, these considerations are rarely incorporated into testing. Given that controllers cannot be optimised for every criterion, a developed controller shall at least be systematically tested after design.

Some theoretical advancements have been made in the process control community, particularly in control performance assessment \cite{srinivasan2012DFAProcess,ding2021Process,dogruer2023process}, yet comprehensive and standardised criteria remain lacking. This need is even more pressing with the rise of AI-based algorithms, which are optimised for specific criteria but require rigorous post-design testing to ensure reliable implementation. From an applied perspective, this paper revisits operational challenges in control systems, proposes standardised testing metrics, and aims to illuminate paths for future improvement. The proposed framework helps prevent premature claims and ensures fair evaluation of new designs.

\section{Operational challenges in control}
\begin{figure}[t!]
    \centerline{\includegraphics[width=0.4\textwidth]{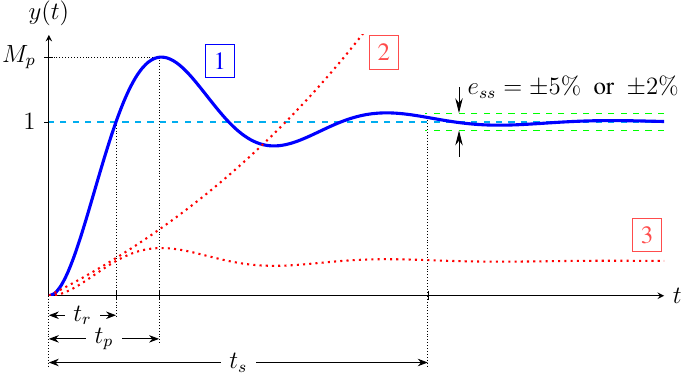}}
        \caption{A typical system step response, 1) stable with acceptable tracking, 2) unstable or divergent 3) stable without tracking}
	\label{fig:typicalstepresponse}
\end{figure}

In the first step, it is crucial to understand the associated challenges with applied control design. As shown in Fig. \ref{fig:realcontrolloop}, a real-world control problem involves a system in a feedback loop, where the controller's main task is to manipulate the system output, $y$, to follow reference signals, $r$, thereby eliminating the error signal, $e$. To achieve this, the controller generates the control signal, $u_c$, which requires an actuator to apply $u_c$ to the system, producing $u_{ac}$. Sensors are used to measure the system output and provide an estimate, $y_m$. A step reference signal is often employed to assess system behaviour, with the system (or environment, in ML terminology) replaced by a model during design stages. A typical response to a step reference is shown in Fig. \ref{fig:typicalstepresponse}, demonstrating the goal of achieving zero tracking error as quickly as possible, quantified through tracking criteria. However, tracking alone is insufficient, as applied control design involves addressing additional key challenges, categorised and explained below:

\begin{figure}[b!]
	\centerline{\includegraphics[width=0.45\textwidth]{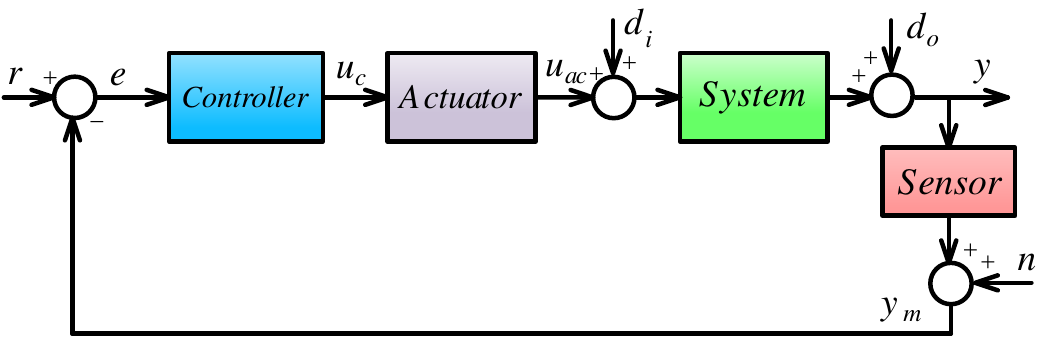}}
        \caption{A real-world feedback control loop}
	\label{fig:realcontrolloop}
\end{figure}

\subsubsection{Actuator and sensor dynamics}
As often ignored in early-stage design but existing subsystems of any control loop, actuators and sensors are integral components of real controllers (Fig. \ref{fig:realcontrolloop}). Sensors introduce various error sources, such as noise, bias, drift, and resolution, which are not easily removed. Similarly, actuators face physical limitations, including the maximum control signal they can handle. Characteristics like amplitude and rate saturation, delay, and phase lag are crucial aspects of actuators that require attention in system design. It is now widely accepted that inadequate consideration of these elements has led to catastrophic and profoundly bitter incidents \cite{SteinRespect}. Therefore, the demanded control signal, $u_c$, should always be feasible for real-life implementation, placing it on par with tracking error—an aspect often ignored in early developments or claims, rendering the algorithm impractical for real-world applications. Sensor and actuator limitations must be considered in the design.

\subsubsection{Disturbance and noise}
\begin{figure}[b!]    \centerline{\includegraphics[width=0.49\textwidth]{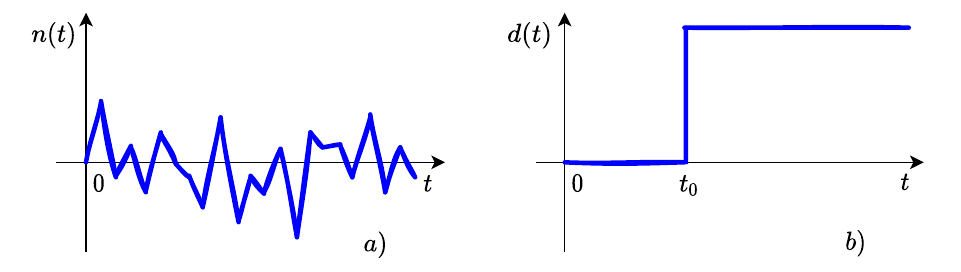}}
        \caption{a) Noise as a stochastic, lower amplitude higher frequency signal, and b) disturbance or a lower frequency higher amplitude signal}
	\label{fig:NoiseandDisturbance}
\end{figure}

Disturbance and noise are two (usually) annoying external signals present in practical systems that need consideration in controller design. 
To be more specific, disturbances can be modelled as external signals entering the loop from system input and output, denoted by $d_i$ and $d_o$. They are signals with low frequency and considerable amplitude and are sometimes called load disturbances. They may appear and disappear after some time, and a small step or rectangular signal can model them. For instance, in a car cruise controller, a bias error in the pedal actuator can be considered as $d_i$, and hitting a speed hump or facing strong wind could be counted as $d_o$. Noise, like disturbances, is another difficulty in a control loop. It is usually a stochastic signal that is added to the sensor output and deteriorates the measurement. Therefore, it is injected after the sensor measurement into the loop. Sometimes noise and disturbance are used interchangeably; however, just to clarify, in control terminology, noise is considered a high-frequency, low-amplitude signal. Noise and disturbance signals are illustrated in Fig. \ref{fig:NoiseandDisturbance}. At times, complete removal of these signals may not be possible, and efforts are made to minimise their adverse effects. Disturbance and noise attenuation are thus crucial properties of any controller. Given their presence at different frequencies, shaping the frequency response of the sensitivity functions of the system is a common approach to address these issues \cite{skogestad2005multivariable,landau2006digital}. In any case, their effects should be, at the very least, considered in time domain simulations during system design and analysis.

\subsubsection{Model uncertainties}
The majority of control techniques utilise a model, to some extent, to design and simulate the system in the early stages. However, dealing with discrepancies between this model and the real system under operational conditions presents a challenge for any controller. Any model is subjected to unknown or uncontrollable error sources that lead to a deviation between the expected and true behaviour. Uncertainty is an estimation of the degree to which the model is imprecise, untrusted, and unknown, and can very simply result from the following factors:

\begin{itemize}
    \item Inaccurate model parameter calculation or estimation.
    \item Variation of parameters over time or changing operating conditions.
    \item High-frequency system dynamics due to simplification or using lower-order models (which is common).
    \item System nonlinearities. 
\end{itemize}

There are various approaches to model uncertainty (\cite{skogestad2005multivariable}), the detailed discussions of which are skipped here. As a simple example, we model the system by the following transfer function:
\vspace{-2pt}
\begin{equation}
    G_p(s) = \frac{K}{\tau s+1}    
\end{equation}

However, the system in practice may behave like one the $G_{p_1}$, $G_{p_2}$, or $G_{p_3}$ as follows:
\vspace{-5pt}
\begin{equation}
\label{eq:uncertainty}
\fontsize{8.95pt}{11pt}\selectfont
    G_{p_1}(s) = \frac{K_1}{\tau_1 s+1}, \; G_{p_2}(s) = \frac{K}{\tau s+1} + \frac{K_2}{\tau_2 s+1}, \; G_{p_3}(s) = \frac{Ke^{-\tau_d}}{\tau s+1}    
\end{equation}

These systems encompass gain/pole location uncertainties, unmodelled dynamics, and delay uncertainties—realistic challenges present in all systems. For instance, the added delay is comparable to the effects of drug or alcohol consumption impairing a person's ability to drive. Various terminologies describe these uncertainties in robust and adaptive control; however, controllers are typically designed to handle only specific categories and ranges of uncertainty. Addressing uncertainties is crucial in practical control design, as designers often encounter unknown-unknown uncertainties. Therefore, a prudent designer ensures sufficient margins, known as robustness. Frequency-domain approaches and metrics such as gain and phase margins serve as fundamental tools for managing uncertainty in linear systems, though their application to nonlinear systems is more complex.

\subsubsection{Non-minimum Phase (NMP) systems}
Another challenge in control design is NMP systems. While classic references exclusively define systems with right-hand plane (RHP) zeros as non-minimum phase, the following systems are considered NMP (\cite{skogestad2005multivariable}):

\begin{itemize}
    \item Systems with unstable (RHP) zeros.
    \item Systems with unstable poles.
    \item Systems with delay.
\end{itemize}
\vspace{-1mm}

Systems with these components exhibit additional phase lag compared to their Minimum Phase counterparts. NMP dynamics drastically limit achievable performance and robustness in feedback control, as discussed in \cite{astromlimitations, SteinRespect, skogestad2005multivariable}, requiring special attention in design. NMP systems with RHP zeros initially move opposite to the desired direction when an input is applied, causing undershoot, a reverse overshoot-like peak. This complicates control, as the system temporarily diverges before stabilising. NMP zeros depend on sensor and actuator selection, seen in bicycle dynamics at low speeds, reversing a car, and aircraft altitude control \cite{NMPzeros}. Undershoot is an inherent characteristic of such systems and cannot be eliminated. However, its peak value can be tuned for high-performance tracking. Reducing response time may increase the undershoot peak, presenting a challenge in dealing with these systems.
In nonlinear systems, the concept of NMP zeros is even more challenging to define than in linear systems, where it relates to the model’s unstable internal dynamics \cite{Sastry1992nonlinearNMPsystems}. Delayed and NMP responses also pose operational challenges in control design that must be addressed to demonstrate a controller’s effectiveness.

We outlined some operational challenges in applied control applications. Humans can learn and adapt to such difficulties; however, existing controllers are system-specific, requiring re-design or re-tuning for each application. Hence, next-generation controllers are expected to provide such versatility, enhancing sustainability in design by preventing resource reallocation. Both theoretical developments and ML-based algorithms should address these challenges in their design and testing. In this paper, we introduce categories of essential design criteria for adequately testing controllers.


\section{Standard Metrics in Applied Control Design}
In the age of LLMs and AI, assume a controller is designed using either ML or rigorous control theory. The key question is: how do we evaluate this controller, and can it reliably transition to real-world use? We treat this controller as a black-box, requiring proper methods for testing and verification. A comprehensive analysis benefits from tailored criteria, regardless of the design objective function, which may cover only a few aspects.

Since feedback relies on the error signal, its convergence remains a key performance measure. However, realising an applied controller requires considering multiple factors beyond error reduction. We categorise standard control metrics into three groups (summarised in Table \ref{table:criteria_summary}):
\begin{enumerate}[label=\textbf{\roman*}), align=left]
    \item Tracking error-based performance metrics.
    \item Energy or control signal-based metrics.
    \item Robustness metrics.
\end{enumerate}

\begin{table}[t!]
\caption{Applied control performance criteria}
\centerline{\includegraphics[width=0.45\textwidth]{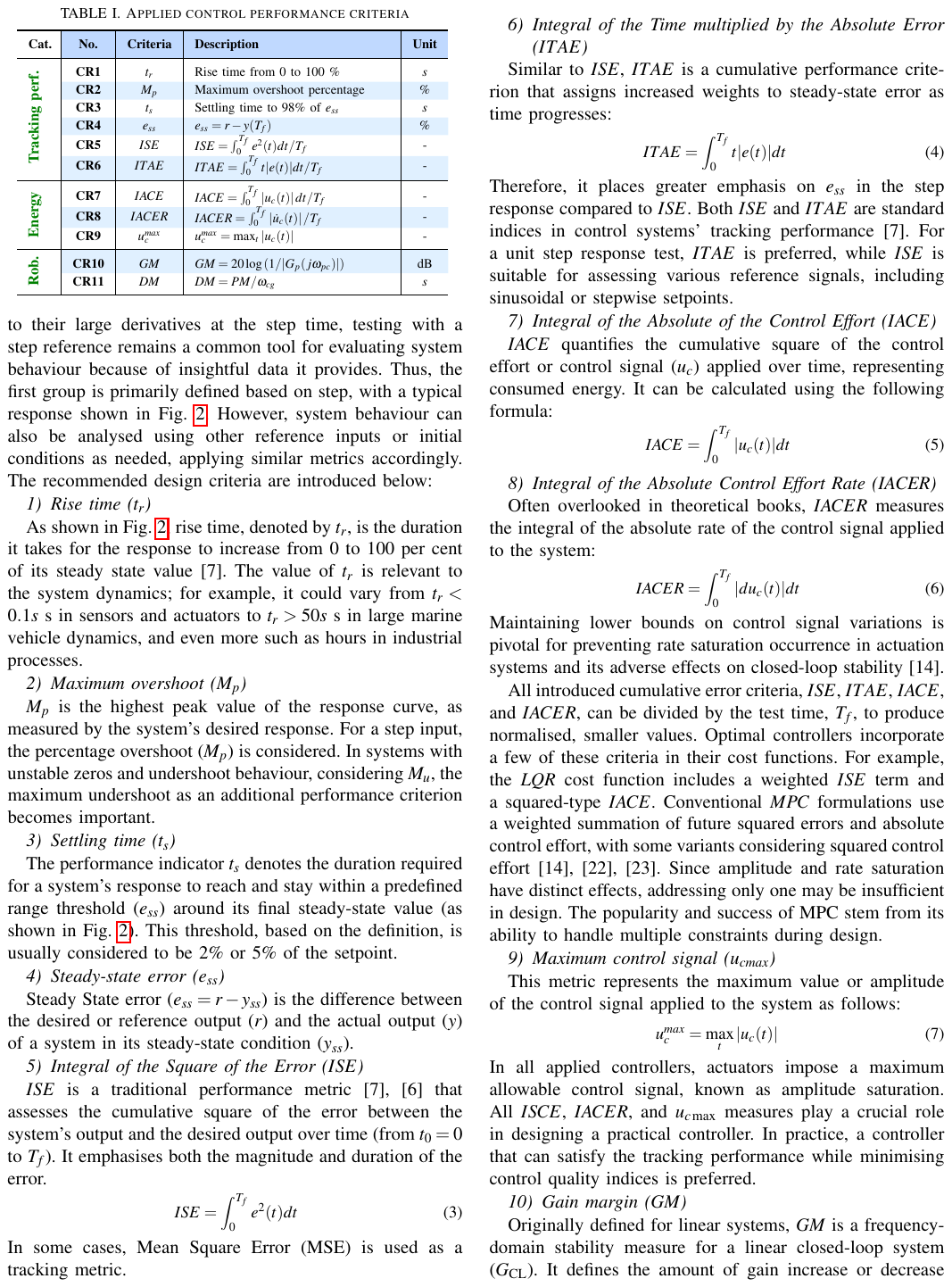}}
\label{table:criteria_summary}
\end{table}

These three groups ensure a proper assessment of any controller providing balance for reasling an applied controller. While step references are rarely used in practice due to their large derivatives at the step time, testing with a step reference remains a common tool for evaluating system behaviour because of insightful data it provides. Thus, the first group is primarily defined based on step, with a typical response shown in Fig. \ref{fig:typicalstepresponse}. However, system behaviour can also be analysed using other reference inputs or initial conditions as needed, applying similar metrics accordingly. The recommended design criteria are introduced below:

\subsubsection{Rise time ($t_r$)}
As shown in Fig. \ref{fig:typicalstepresponse}, rise time, denoted by $t_r$, is the duration it takes for the response to increase from $0$ to $100$ per cent of its steady state value \cite{levine2018control}. The value of $t_r$ is relevant to the system dynamics; for example, it could vary from $t_r < 0.1s$ s in sensors and actuators to $t_r > 50s$ s in large marine vehicle dynamics, and even more such as hours in industrial processes.

\subsubsection{Maximum overshoot ($M_p$)}
$M_p$ is the highest peak value of the response curve, as measured by the system's desired response. For a step input, the percentage overshoot ($M_p$) is considered. In systems with unstable zeros and undershoot behaviour, considering $M_u$, the maximum undershoot as an additional performance criterion becomes important.

\subsubsection{Settling time ($t_s$)}
The performance indicator $t_s$ denotes the duration required for a system's response to reach and stay within a predefined range threshold ($e_{ss}$) around its final steady-state value (as shown in Fig. \ref{fig:typicalstepresponse}). This threshold, based on the definition, is usually considered to be $2\%$ or $5\%$ of the setpoint.

\subsubsection{Steady-state error ($e_{ss}$)}
Steady State error ($e_{ss} = r - y_{ss}$) is the difference between the desired or reference output ($r$) and the actual output ($y$) of a system in its steady-state condition ($y_{ss}$).

\subsubsection{Integral of the Square of the Error ($ISE$)}
$ISE$ is a traditional performance metric \cite{levine2018control,ControlPerformance1961} that assesses the cumulative square of the error between the system's output and the desired output over time (from $t_0=0$ to $T_f$). It emphasises both the magnitude and duration of the error.
\begin{equation}\
ISE=\int_0^{T_f} e^2(t) d t \
\end{equation}
In some cases, Mean Square Error (MSE) is used as a tracking metric.
\subsubsection{Integral of the Time multiplied by the Absolute Error ($ITAE$)}
Similar to $ISE$, $ITAE$ is a cumulative performance criterion that assigns increased weights to steady-state error as time progresses:
\begin{equation}
ITAE=\int_0^{T_f} t|e(t)| d t \
\end{equation}
Therefore, it places greater emphasis on $e_{ss}$ in the step response compared to $ISE$. Both $ISE$ and $ITAE$ are standard indices in control systems’ tracking performance \cite{levine2018control}. For a unit step response test, $ITAE$ is preferred, while $ISE$ is suitable for assessing various reference signals, including sinusoidal or stepwise setpoints.

\subsubsection{Integral of the Absolute of the Control Effort ($IACE$)}
$IACE$ 	quantifies the cumulative absolute of the control effort or control signal ($u_c$) applied over time, representing consumed energy. It can be calculated using the following formula:
\begin{equation}\
IACE=\int_0^{T_f}|u_c(t)| d t \
\end{equation}

\subsubsection{Integral of the Absolute Control Effort Rate (IACER)}
Often overlooked in theoretical books, $IACER$ measures the integral of the absolute rate of the control signal applied to the system:
\begin{equation}\
IACER=\int_0^{T_f}|d u_c(t)| dt \
\end{equation}
Maintaining lower bounds on control signal variations is pivotal for preventing rate saturation occurrence in actuation systems and its adverse effects on closed-loop stability \cite{SteinRespect}. 

All introduced cumulative error criteria, $ISE$, $ITAE$, $IACE$, and $IACER$, can be divided by the test time, $T_f$, to produce normalised, smaller values. Optimal controllers incorporate a few of these criteria in their cost functions. For example, the $LQR$ cost function includes a weighted $ISE$ term and a squared-type $IACE$. Conventional $MPC$ formulations use a weighted summation of future squared errors and absolute control effort, with some variants considering squared control effort \cite{SteinRespect,duda1997ratesaturation,turner2020ratesatur}. Since amplitude and rate saturation have distinct effects, addressing only one may be insufficient in design. The popularity and success of MPC stem from its ability to handle multiple constraints during design.

\subsubsection{Maximum control signal ($u_{cmax}$)}
This metric represents the maximum value or amplitude of the control signal applied to the system as follows:
\begin{equation}
u_{c}^{max} =\max _t\left|u_c(t)\right|
\end{equation}
In all applied controllers, actuators impose a maximum allowable control signal, known as amplitude saturation. All $ISCE$, $IACER$, and $u_{c \max }$ measures play a crucial role in designing a practical controller. In practice, a controller that can satisfy the tracking performance while minimising control quality indices is preferred. 
\subsubsection{Gain margin ($GM$)}
Originally defined for linear systems, $GM$ is a frequency-domain stability measure for a linear closed-loop system ($G_{\text{CL}}$). It defines the amount of gain increase or decrease in the open-loop system ($K_1$ in \ref{eq:uncertainty} $G_{P_1}$)  before the system becomes unstable. Typically, $GM > 6\; \text{dB}$ and $GM < -6\; \text{dB}$ (or equivalently, $GM > 2$ and $GM < 0.5$) are considered desired values; however, this can vary depending on the system requirements. In theoretical textbooks, only the maximum value of $GM$ is considered, while its minimum value is also important for conditionally stable systems. Although $GM$ is primarily defined for linear systems, experienced control engineers apply such $+6\; \text{dB}$ or even higher tests at operating points of their nonlinear systems, and in Hardware-in-the-Loop (HIL) simulations, as an acceptance test for controllers and to identify potential weaknesses. This is feasible because well-designed systems rarely exhibit strange or extremely chaotic behaviour in most engineering applications. The importance of estimating frequency-domain robustness margins for black-box systems is well recognised in the literature and continues to be investigated in greater theoretical depth \cite{isoshima2023GMPMdatadriven}.

\subsubsection{Delay Margin ($DM$)}
Phase margin ($PM$) is another robustness metric, but its application to nonlinear systems can be difficult to test. However, as a robustness metric that can be easily applied for testing, delay margin ($DM$) defines the maximum delay that can be added to the system while maintaining stability. 

It is known that a system with acceptable $GM$ and $PM$ can exhibit poor $DM$ \cite{landau2006digital}. The ECSS handbook also emphasises this metric \cite{ECSS2010}. Similar to $GM$, $DM$ can also be applied in testing of any control loop, by artificially increasing the delay in the system’s input. Some classical textbooks consider the minimum value of the delay margin to be greater than the system's sampling time ($DM > T_s$) \cite{landau2006digital}. However, given today's small sampling rates, this should be increased to a reasonable amount depending on the system's dynamics, which can be around $10\%$ of the system's rise time.

These criteria are summarised in Table \ref{table:criteria_summary}. To generate normalised values for cumulative criteria, a division by $T_f$ (simulation time) is suggested to avoid excessively large numbers. In any controller design, one or more of the above criteria are typically used. However, to ensure an effective controller, metrics from all three categories (i-iii) should be considered. In practice, this is the approach typically adopted by experienced industrial engineers. Depending on the problem, researchers may choose additional measures such as maximum undershoot, bandwidth, modulus margin, or, in MIMO systems, sensitivity functions. This applies to nonlinear systems in practical scenarios, though less so in purely theoretical problems. The next section presents an appealing example of testing various controllers using these metrics.
\section{Simulation studies}
The plant under control is an empirical model of the Bounded Input Bounded Output (BIBO) unstable yaw motion of the REMUS Autonomous Underwater Vehicle (AUV) \cite{prestero2001verification,sarhadi2016model}, described by the following state-space equations:
\begin{equation}
\begin{aligned}
\dot{x} &= \begin{bmatrix} 
\dot{v} \\ 
\dot{r} \\ 
\dot{\psi} 
\end{bmatrix}
= \begin{bmatrix} 
-2.72 & -0.43 & 0 \\ 
-3.38 & -2.51 & 0 \\ 
0 & 1 & 0 
\end{bmatrix} 
\begin{bmatrix} 
v \\ 
r \\ 
\psi 
\end{bmatrix}
+ \begin{bmatrix} 
0.24 \\ 
-1.82 \\ 
0 
\end{bmatrix} \delta_r, \\
y &= \begin{bmatrix} 
0 & 0 & 1
\end{bmatrix} x.
\end{aligned}
\end{equation}

The aim is to control the vehicle's yaw motion ($\psi$ angle) using the rudder actuation input ($\delta_r$). Two tests are defined: \textbf{\small{T1})} a nominal step response test in the linear operating region, and \textbf{\small{T2})} a $20^\circ$ large step response test in the nonlinear operating domain with input saturation of $|u_c| \leq 20^\circ$, rate saturation of $|\dot{u}_c| \leq 20^\circ$, and a multiplicative unmodeled dynamic of $G_u(s) = \frac{225}{s^2 + 12s + 225}$, which represents a worst-case scenario compared to Rohr Test's (RT) example. Moreover, an output disturbance of $d_o = 2$ and noise with a standard deviation of $\sigma = 0.05$ are introduced at the $15^{\text{th}}$ and $20^{\text{th}}$ seconds, respectively. Six controllers, a combination of classical and novel approaches, i.e., $C_1 - C_6$, are designed, where $C_5$ and $C_6$ are modified versions of $C_1$ and $C_3$, respectively, to address input constraints (only activated in the second test). These controllers will be revealed after presenting the results, maintaining the paper's black-box testing approach.
\begin{figure}[t]
    \centering    \centerline{\includegraphics[width=0.49\textwidth]{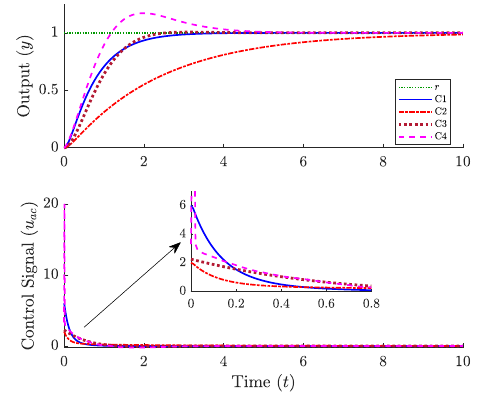}}
    \caption{Test 1 results for various controllers: nominal condition}
    \label{fig:Test1}
\end{figure}
Fig. \ref{fig:Test1} depicts the step response results under nominal conditions, with the corresponding criteria summarised in Table \ref{table:Test1}. Tables use colour codes to highlight worse metrics in red and the best in green. According to these results, all controllers successfully perform the tracking, with C1 and C3 demonstrating the fastest responses, closely matching each other. However, C3 exhibits better control indices, particularly in control rate. C3 shows notably higher $DM$ and lower $GM$, while C1 has a poorer delay margin. C4 offers a fast response in terms of $ISE$, but not $ITAE$, at the cost of higher control energy. Although C1 and C3 show the best overall responses, C2, while exhibiting good energy metrics and $GM$, is the slowest controller and may not be considered due to its slow response. Therefore, C1 and C3 emerge as viable options, although further tests are required.

\begin{table}[t!]
\caption{Performance Comparison for Test 1}
\centerline{\includegraphics[width=0.4\textwidth]{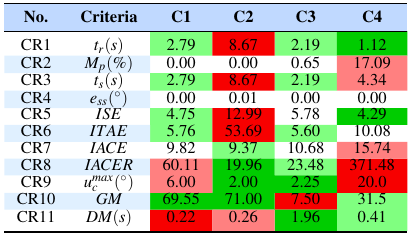}}
\label{table:Test1}
\end{table}

\begin{figure}[b]
    \centering    \centerline{\includegraphics[width=0.5\textwidth]{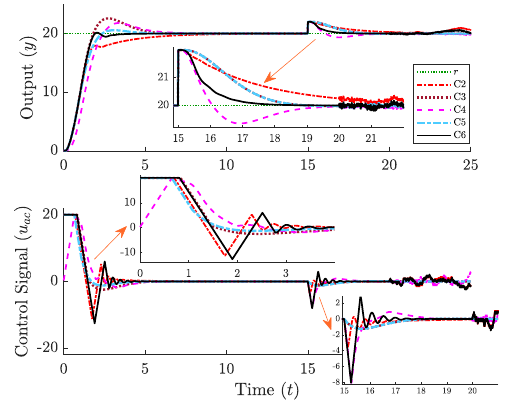}}
    \caption{Test 2 results considering saturations, uncertainty, and perturbations}
    \label{fig:Test2}
\end{figure}
The second test is particularly challenging due to the operation in the nonlinear domain, where amplitude and rate saturations, unmodeled uncertainty (RT), disturbances, and noise are present. Fig. \ref{fig:Test2} presents the results, with the corresponding criteria summarised in Table \ref{table:Test2}. Surprisingly, in this test, C1, despite its appealing performance in Test 1, becomes unstable, so its results are not shown. However, we can observe that C5 and C6 exhibit better performance, while C3 and C4 also demonstrate acceptable performance, even under these challenging conditions. Based solely on Fig. \ref{fig:Test2}, one might choose C6 as the best controller. However, referring to Table \ref{table:Test2}, C5 demonstrates significantly lower control effort and better robustness, especially in terms of delay tolerance. Ultimately, controller selection is typically based on the overall system requirements.

\begin{table}[b!]
\caption{Performance Comparison for Test 2}
\centerline{\includegraphics[width=0.45\textwidth]{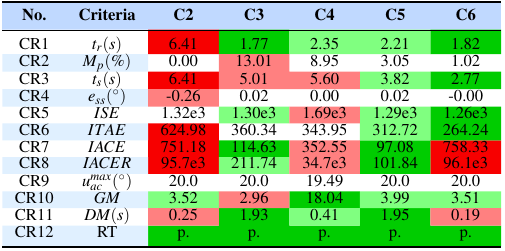}}
\label{table:Test2}
\end{table}

\section{Discusion}
Now, we disclose the controllers. C1 (PID1) was a parallel Proportional-Derivative (PD) controller with $K_p = 6$ and $K_d = 4$. While C1 exhibited the best performance in the first test, it struggled to handle perturbations. C6 (PID-AW) was a modified version of C1 with an anti-windup gain of $K_{aw} = 4$. The results show how only an additional parameter can boost the controller's performance, enabling it to better handle real-life operational challenges. C2 (PID2) was a low-gain PD with $K_p = 2$ and $K_d = 4$, offering a slower response but decent robustness. C3 (MPC1) was a non-constrained MPC with a prediction horizon of $N_y = 120$, a control horizon of $N_u = 1$, and a control rate weight of $\lambda = 0.1$. C5 (MPC2) was a constrained MPC that accounted for both amplitude and rate saturation levels ($|u_c| < \pm 20$ and $|\dot{u}_c| < \pm 30$).
Both MPCs, C3 and C5, performed well in the challenging Test 2, demonstrating very smooth control signal performance metrics and a significantly larger delay margin. Of course, the performance of the constrained MPC, C5, was the best. C4 was a DDPG controller \cite{lillicrap2015continuousddpg} tuned with a linear quadratic reward function, with $Q=1$ for tracking and $R=0.01$ for the control signal. C4 showed acceptable nonlinear performance with some adaptation in response. For instance, the maximum control signal in T1 was high, but DDPG was able to adapt to the new situation, which was not expected from linear controllers. It also demonstrated excellent tolerance to gain uncertainty in T2 and decent delay tolerance. It should be noted that DDPG is a data-driven approach and does not benefit from a model but requires a cumbersome tuning process (on the order of several days). The DDPG was implemented in Python using PyTorch, while the other controllers were simulated using MATLAB, with results shared on GitHub for further details \footnote{\href{https://github.com/Psarhadi/Performance_Criteria_Control_PID_MPC_ML}{\textcolor{blue}{github.com/Psarhadi/Performance\_Criteria\_Control\_PID\_MPC\_ML}}}. It is worth mentioning that there was an intention to also test Model-Free Adaptive Control (MFAC), one of the latest developments in data-driven control \cite{hou2019MFACFFDLDatadriven,khaki2023datadriven}. Acknowledging that MFAC was originally developed for BIBO-stable systems, a proportional controller was used to stabilise the system, but satisfactory results were not achieved with Full-Form Dynamic Linearisation (FFDL), MFAC's most prominent approach \cite{hou2019MFACFFDLDatadriven}. This may be due to limitations in the author's understanding and is open to reconsideration; however, a similar issue was reported in \cite{narendra2024ML}.

The PID controller with anti-windup (C6) and constrained MPC (C5) demonstrated better performance, with MPC being superior. However, it should be noted that PID achieves this performance with only three parameters and a very straightforward tuning procedure and implementation requirements. This simplicity makes PID popular for industrial applications, as it is easy to design, re-tune and implement. The success of PID in practice is well-documented \cite{hagglund2024PIDMPC}, and its effectiveness has been further proven through modifications \cite{skogestad2023advanced}. These results reaffirm PID’s practical effectiveness, while MPC and ML-based controllers can handle higher levels of complexity and decision-making loops \cite{hagglund2024PIDMPC,Hadi2024JOE}, incorporating more intricate situations in their optimisation framework. To clarify, ML-based controllers are not intended to replace classical controllers for trivially solvable problems but to address challenges that are difficult to model or understand, such as decision-making tasks. ML-based controllers are not intended to replace classical controllers for trivial problems that are already solved with classical controllers. Instead, they can tackle challenges that are hard to model or understand, such as complex decision-making tasks. They can also enhance adaptability beyond classical controllers, which typically suit only specific system categories. This has already occurred in image, voice, and text processing but is yet to be realised in control systems applications. Overall, this paper aims to establish rigorous, fair, and standardised procedures for verifying control algorithms in applied problems. 

\section{CONCLUSION}
Utilising a systematic and big-picture view and building on many years of advancements in control literature, which are well-known but sometimes overlooked, a framework and standard test criteria were introduced to evaluate the viability of controllers for applied problems. After introducing the operational challenges in applied control systems, the paper provides a solid foundation of standard test criteria to test and verify any controller, ensuring a bias-free comparison. It is demonstrated that a controller that may behave acceptably under nominal conditions can become ineffective when faced with real-world challenges. The paper also briefly performed a rigorous comparison between several controllers, including PID, MPC, and a well-known ML technique. The goal is to utilise such comprehensive methods in system design to prevent redesigns and ensure sustainability. 
An important takeaway is the emphasis on testing any controller across three categories of criteria: tracking, control energy, and robustness, even if the controller is theoretically proven for only one or two of these criteria. It should not be ignored that professional engineers routinely perform advanced tests and statistical analyses using tools such as Monte Carlo simulations or Software/Hardware-in-the-Loop (SIL/HIL) to verify controller applicability during system design. Although the exploitation of criteria may vary depending on the problem, or extra tests may be required to analyse a system, the aforementioned metrics are fundamental for quantifying system behaviour. Similar metrics are also applicable for the evaluation of complex systems, such as algorithms used in autonomous vehicles, where precise performance evaluations are crucial. The main limitation was another challenge in applied control: the real-time implementability of algorithms, which can be addressed using tools such as SIL or HIL. A clear understanding of these metrics can significantly enhance system design, leading to more reliable and efficient outcomes in real-world applications.

\bibliographystyle{IEEEtran}
\bibliography{cdcconf.bib}

\begin{thebibliography}{10}
\providecommand{\url}[1]{#1}
\csname url@samestyle\endcsname
\providecommand{\newblock}{\relax}
\providecommand{\bibinfo}[2]{#2}
\providecommand{\BIBentrySTDinterwordspacing}{\spaceskip=0pt\relax}
\providecommand{\BIBentryALTinterwordstretchfactor}{4}
\providecommand{\BIBentryALTinterwordspacing}{\spaceskip=\fontdimen2\font plus
\BIBentryALTinterwordstretchfactor\fontdimen3\font minus \fontdimen4\font\relax}
\providecommand{\BIBforeignlanguage}[2]{{%
\expandafter\ifx\csname l@#1\endcsname\relax
\typeout{** WARNING: IEEEtran.bst: No hyphenation pattern has been}%
\typeout{** loaded for the language `#1'. Using the pattern for}%
\typeout{** the default language instead.}%
\else
\language=\csname l@#1\endcsname
\fi
#2}}
\providecommand{\BIBdecl}{\relax}
\BIBdecl

\bibitem{skogestad2023advanced}
S.~Skogestad, ``Advanced control using decomposition and simple elements,'' \emph{Annual Reviews in Control}, vol.~56, p. 100903, 2023.

\bibitem{hagglund2024PIDMPC}
T.~H{\"a}gglund and J.~L. Guzm{\'a}n, ``Give us {PID} controllers and we can control the world,'' \emph{IFAC-OnLine}, vol.~58, no.~7, pp. 103--108, 2024.

\bibitem{lillicrap2015continuousddpg}
T.~P. Lillicrap, J.~J. Hunt, A.~Pritzel, N.~Heess, T.~Erez, Y.~Tassa, D.~Silver, and D.~Wierstra, ``Continuous control with deep reinforcement learning,'' \emph{arXiv preprint arXiv:1509.02971}, 2015.

\bibitem{skogestad2019smith}
C.~Grimholt and S.~Skogestad, ``Should we forget the {Smith Predictor?}'' 2019.

\bibitem{narendra2024ML}
K.~S. Narendra and K.~George, ``Can model-free controllers for complex systems stabilize and provide satisfactory response?'' in \emph{2024 American Control Conference (ACC)}.\hskip 1em plus 0.5em minus 0.4em\relax IEEE, 2024, pp. 1237--1242.

\bibitem{ControlPerformance1961}
W.~Schultz and V.~Rideout, ``Control system performance measures: Past, present, and future,'' \emph{IRE transactions on automatic control}, no.~1, pp. 22--35, 1961.

\bibitem{levine2018control}
W.~S. Levine, \emph{The Control Handbook}.\hskip 1em plus 0.5em minus 0.4em\relax CRC press, 2018.

\bibitem{skogestad2005multivariable}
S.~Skogestad and I.~Postlethwaite, \emph{Multivariable feedback control: analysis and design}.\hskip 1em plus 0.5em minus 0.4em\relax john Wiley \& sons, 2005.

\bibitem{dydek2010adaptiveX15}
Z.~T. Dydek, A.~M. Annaswamy, and E.~Lavretsky, ``Adaptive control and the nasa {X-15-3} flight revisited,'' \emph{IEEE Control Systems Magazine}, vol.~30, no.~3, pp. 32--48, 2010.

\bibitem{rohrs1985robustness}
C.~Rohrs, L.~Valavani, M.~Athans, and G.~Stein, ``Robustness of continuous-time adaptive control algorithms in the presence of unmodeled dynamics,'' \emph{IEEE Transactions on Automatic Control}, vol.~30, no.~9, pp. 881--889, 1985.

\bibitem{ECSS2010}
{European Cooperation on Space Standardisation}, \emph{{Control Performance Guideline Handbook ECSS-E-ST-60-10A}}, ESA-ESTEC Requirements \& Standards Division, 2010.

\bibitem{wie1992benchmark}
B.~Wie and D.~S. Bernstein, ``Benchmark problems for robust control design,'' \emph{Journal of Guidance, Control, and Dynamics}, vol.~15, no.~5, pp. 1057--1059, 1992.

\bibitem{maestre2025benchmark2}
J.~M. Maestre and C.~Ocampo-Martinez, \emph{Control Systems Benchmarks}.\hskip 1em plus 0.5em minus 0.4em\relax Springer, 2025.

\bibitem{SteinRespect}
G.~Stein, ``Respect the unstable,'' \emph{IEEE Control Systems Magazine}, vol.~23, no.~4, pp. 12--25, 2003.

\bibitem{srinivasan2012DFAProcess}
B.~Srinivasan, T.~Spinner, and R.~Rengaswamy, ``Control loop performance assessment using detrended fluctuation analysis {(DFA)},'' \emph{Automatica}, vol.~48, no.~7, pp. 1359--1363, 2012.

\bibitem{ding2021Process}
S.~X. Ding and L.~Li, ``Control performance monitoring and degradation recovery in automatic control systems: A review,'' \emph{Control Engineering Practice}, vol. 111, p. 104790, 2021.

\bibitem{dogruer2023process}
T.~Dogruer, ``Design of {I-PD} controller based modified smith predictor for processes with inverse response and time delay using equilibrium optimizer,'' \emph{IEEE Access}, vol.~11, pp. 14\,636--14\,646, 2023.

\bibitem{landau2006digital}
I.~D. Landau and G.~Zito, \emph{Digital control systems: design, identification and implementation}.\hskip 1em plus 0.5em minus 0.4em\relax Springer, 2006, vol. 130.

\bibitem{astromlimitations}
K.~J. Astrom, ``Limitations on control system performance,'' \emph{European Journal of Control}, vol.~6, no.~1, pp. 2--20, 2000.

\bibitem{NMPzeros}
J.~B. Hoagg and D.~S. Bernstein, ``Nonminimum-phase zeros - much to do about nothing - classical control - revisited part {II},'' \emph{IEEE Control Systems Magazine}, vol.~27, no.~3, pp. 45--57, 2007.

\bibitem{Sastry1992nonlinearNMPsystems}
J.~Hauser, S.~Sastry, and G.~Meyer, ``Nonlinear control design for slightly non-minimum phase systems: Application to {V/STOL} aircraft,'' \emph{Automatica}, vol.~28, no.~4, pp. 665--679, 1992.

\bibitem{duda1997ratesaturation}
H.~Duda, ``Prediction of pilot-in-the-loop oscillations due to rate saturation,'' \emph{Journal of Guidance, Control, and Dynamics}, vol.~20, no.~3, pp. 581--587, 1997.

\bibitem{turner2020ratesatur}
M.~C. Turner, J.~Sofrony, and E.~Prempain, ``Anti-windup for model-reference adaptive control schemes with rate-limits,'' \emph{Systems \& Control Letters}, vol. 137, p. 104630, 2020.

\bibitem{isoshima2023GMPMdatadriven}
K.~Isoshima, M.~Tanemura, and Y.~Chida, ``Data-driven estimation of the lower bounds of gain and phase margins,'' \emph{Automatica}, vol. 153, p. 111008, 2023.

\bibitem{prestero2001verification}
T.~T.~J. Prestero, ``Verification of a six-degree of freedom simulation model for the {REMUS} autonomous underwater vehicle,'' Ph.D. dissertation, Massachusetts institute of technology, 2001.

\bibitem{sarhadi2016model}
P.~Sarhadi, A.~R. Noei, and A.~Khosravi, ``Model reference adaptive {PID} control with anti-windup compensator for an autonomous underwater vehicle,'' \emph{Robotics and Autonomous Systems}, vol.~83, pp. 87--93, 2016.

\bibitem{hou2019MFACFFDLDatadriven}
Z.~Hou and S.~Xiong, ``On model-free adaptive control and its stability analysis,'' \emph{IEEE Transactions on Automatic Control}, vol.~64, no.~11, pp. 4555--4569, 2019.

\bibitem{khaki2023datadriven}
A.~Khaki-Sedigh, \emph{An Introduction to Data-Driven Control Systems}.\hskip 1em plus 0.5em minus 0.4em\relax John Wiley \& Sons, 2023.

\bibitem{Hadi2024JOE}
B.~Hadi, A.~Khosravi, and P.~Sarhadi, ``Adaptive formation motion planning and control of autonomous underwater vehicles using deep reinforcement learning,'' \emph{IEEE Journal of Oceanic Engineering}, vol.~49, no.~1, pp. 311--328, 2024.

\end{thebibliography}
\end{document}